\documentclass[twocolumn]{NobArticle}
\runninghead{Shortened Running Article Title}
\footertext{\textit{Journal X} (2023) 12:684}
\usepackage{graphicx} % Add in preamble
\usepackage{tabularx}
\title{ZeroML: A Next Generation AutoML Language}

\author{
    Monirul Islam Mahmud\textsuperscript{1} \\
    \textsuperscript{\textbf{1}}
    Department of Computer and Information Science, Fordham University
}

\usepackage{tabularx}

\renewcommand{\maketitlehookd}{%
\begin{abstract}
    \noindent ZeroML is a new generation programming language for AutoML to drive the ML pipeline in a compiled and multi-paradigm way, with a pure functional core. Meeting the shortcomings introduced by Python, R, or Julia such as slow-running time, brittle pipelines or high dependency cost ZeroML brings the Microservices-based architecture adding the modular, reusable pieces such as DataCleaner, FeatureEngineer or ModelSelector. As a native multithread and memory-aware search optimized toolkit, and with one command deployability ability, ZeroML ensures non-coders and ML professionals to create high-accuracy models super fast and in a more reproducible way. The verbosity of the language ensures that when it comes to dropping into the backend, the code we’ll be creating is extremely clear but the level of repetition and boilerplate required when developing on the front end is now removed.
    
    \medskip

    \small{\textbf{Index Terms:} ZeroML, Programming Language, AutoML.}
\end{abstract}
}

\begin{document}

\small
\maketitle

% Introduction
\section{Introduction}

AutoML (Automated Machine Learning) has evolved as an area of significant research signifying the bourgeoning of AI era in simplifying ML tasks such as data preprocessing, feature creation, model selection, hyperparameter tuning. In the previous years, new flavors for popular AutoML frameworks have appeared with increased abilities (e.g. Auto-sklearn 2.0, TPOT, H2O-3, AutoKeras, FLAML, Google’s Vizier), frequently based upon meta-learning or multi-objective optimization to increase efficiency and accuracy over large-scale problems. Figure: A standard AutoML pipeline (data processing, feature engineering, model production, model evaluation). Typical AutoML process includes data collection and preprocessing, feature selection / extraction, mod- el search (hypothesis class such as SVMs, random forests, CNNs, and also their hyperparameters or architectures), and model evaluation. You will find that the latter steps are automated end-to-end for you in a modern framework. For instance, Auto-sklearn 2.0 adopts a straightforward meta-feature-free meta-learning scheme and bandit-budgeting to facilitate hands-free AutoML for large datasets that are severely time-constrained. This next-wave AutoML advances the current state-of-the-art by orders of magnitude—from hours to seconds of computation time on many benchmarks. Another source for training strategy (now available as open-source in Google Cloud), Google Vizier offers a scalable black-box optimization service for hyperparameter tuning and transfer learning for studies. Other lightweight libraries, such as Microsoft’s FLAML, prioritize cost: FLAML orders trials by cheapest to most expensive, but does so by going towards a more accurate model, removing the need for the data scientist to look for the last useful model while not wasting compute[2,3]. Some other popular tools include TPOT (Genetic-programming hybridized with Natural Language Processing through an improved interface to symbolic regression), H2O AutoML (open-source/enterprise tool for Automatic Machine Learning, exploratory autoML suite for deep learning), AutoKeras (Neural Architecture Search for deep learning), and PyCaret or AutoGluon (easy-to-use high-level machine learning library).

The standard process in ML is divided into 5 main stages – define problem, acquire data, process/prepare your data, use your data to train your model, and iterate upon your model with evaluation as the common feedback. Each of these processes is tedious and consume time based on human brain and expertise. For example, in the field of ML and AI, data preprocessing procedures include cleaning, normalization, handling missing values, encoding, model training requires thoughtful algorithm selection, manual hyperparameter tuning, and iterative validation. By comparison, the AutoML workflow eases the entire workflow by automate the crucial steps including the preprocessing, model selection, training and evaluation. Once the problem has been described and the data gathered, the user can turn the rest over to AutoML tools. These tools do the ETL, feature engineering, algorithm and model tuning internally, saving a significant amount of time and technical knowhow. This automation allows novice users in machine learning, like business analysts, researchers, and students, to produce high-quality models without knowing much about the internal mechanisms.

The efficiency of AutoML is very useful for fast prototyping and large scale use. It moves the user's attention away from low-level engineering towards high-level thinking, like making sense of model outputs and incorporating the results into a business process or scientific workflow. Traditional AutoML frameworks are written in high-level languages, such as Python or R, that have rich libraries (scikit-learn, TensorFlow, H2O, etc.) while, at the same time, have some constraints. Due to Python's Global Interpreter Lock (GIL) and dynamic typing, pure-Python loops are slow (and the only way to get around that is usually to replace them with C/C++/CUDA extensions) which prevents scalig to multicore. Even Python AutoML toolchains could break due to dependencies across OS and versions (e.g., Auto-sklearn has historically been Windows unsupported, limiting a portability across different platforms and different Python versions.) Another issue \& memory inefficiency: iteratively evaluating numerous large models is quite taxing on RAM and almost always requires specialized memory management. R has good statistical tooling (e.g. the caret package, H2O’s R interface) but even fewer AutoML tools and similar scaling limitations. Having new abstractions is relevant because current ML languages were not designed for hyperparameter search or pipeline composition. Domain-specific languages (DSLs) or pipeline compilers (outside of Jupyter notebooks) have been proposed as a way to more explicitly represent and optimize AutoML workflows (including the use of graph IRs like ONNX/MLIR).

The current AutoML libraries have pointed out a few pain points in the literature. For instance, a 2023 review of TPOT found that under resource-limited conditions (such as memory and timeout errors), it could conduct only 43\% of the search tasks. Although PyCaret has superior functionality and is very fast, it had a lesser model accuracy and less customizable. R’s AutoML tools – such as Caret and H2O – require complicated library integrations, and have a less robust parallelism. Furthermore, deployment in the majority of existing systems is manual, error-prone and time-consuming, often involving retraining models or re-building environments. We aim to fill these gaps with the velocity signatures, through the following key contributions of ZeroML: 

\begin{itemize}
	\item AutoML have natively built in Data preprocessing, model search, evaluation and deployment is all part of the language syntax included in the core language without the need of external packages.
	\item It is faster and more parallelized than python and R since the former two are considered interpreted.
	
	\item It allows the developers to independently evolve objects such as DataCleaner, FeatureEngineer, ModelSelector making debugging and experimentation simpler. Deploy to API, edge, or serverless with 1 line; NO re-training.
\end{itemize}

\section{Target Domains \& Users}
ZeroML is designed to serve a diverse set of users, including business analysts, domain experts, students, engineers, and researchers. Business users can run credit scoring or sales forecasting without deep coding skills. Scientists and marketers can build models quickly with minimal syntax. Students benefit from its simplicity to focus on learning core ML concepts. Engineers can prototype and deploy pipelines faster due to ZeroML’s compiled nature and modular structure. Researchers can test hypotheses efficiently using customizable AutoML components. Overall, ZeroML bridges the gap between ML complexity and practical usability, offering speed, readability, and deployment readiness for all user levels.

\section{Type of Language}

\subsection{Compiled Language}
ZeroML is a compiled, multi-paradigm programming language specifically designed to address the performance and modularity limitations of existing AutoML tools. Being a compiled language, ZeroML translates source code into optimized machine code before execution, resulting in significantly faster runtime performance than interpreted languages such as Python or R. This also allows ZeroML to leverage advanced multithreading and native parallelization, making it highly efficient for large-scale searches and model training tasks. Additionally, the compiled nature helps minimize runtime errors and offers better memory management, a crucial feature when working with large datasets in real-world machine learning applications.
\subsection{Multi Paradigm Programming}
ZeroML also adopts a multi-paradigm programming style, offering both procedural and object-oriented constructs to accommodate users with varying levels of expertise. For beginners or users seeking simplicity, ZeroML supports procedural programming where workflows follow a clear, ordered sequence: Load → Search → Deploy. This linear, readable structure allows domain experts and students to quickly grasp the pipeline without needing to understand complex software engineering patterns. For advanced users, ZeroML introduces an object-oriented design, where key components of the AutoML process—such as DataCleaner, FeatureEngineer, ModelSelector, Evaluator, and Deployer—are encapsulated as modular objects. This promotes reusability, separation of concerns, and flexibility in customizing individual parts of the pipeline. Users can extend or override specific behaviors (e.g., writing a custom evaluator or optimizer) without modifying the entire workflow.

\begin{table}[h!]
\centering
\small
\resizebox{\columnwidth}{!}{%
\begin{tabular}{|p{2.8cm}|p{6.2cm}|}
\hline
\textbf{Readability} & \textbf{Moderate.} Still requires learning new syntax \& patterns, making it slightly more complex than beginner Python. \\
\hline
\textbf{Writability} & \textbf{High.} ZeroML workflows can be written in very few lines compared to Python or R. \\
\hline
\textbf{Reliability} & \textbf{Moderate.} ZeroML has no external library dependency, no environment mismatch errors. However, it offers less support for custom transformers or hybrid models. \\
\hline
\textbf{Runtime Cost} & \textbf{Low.} Compiled language runs faster than interpreted languages like Python and R. \\
\hline
\end{tabular}
}
\caption{Comparison of ZeroML with Python and R}
\label{tab:zeroml}
\end{table}

This hybrid design ensures that ZeroML is both approachable and extensible. Beginners can write functional AutoML code in a few lines, while experienced developers have full control over pipeline internals. This balance between simplicity and power makes ZeroML a strong candidate for becoming the next-generation AutoML-focused language.

\begin{figure}[!htpb]
    \centering
    \includegraphics[width=\linewidth]{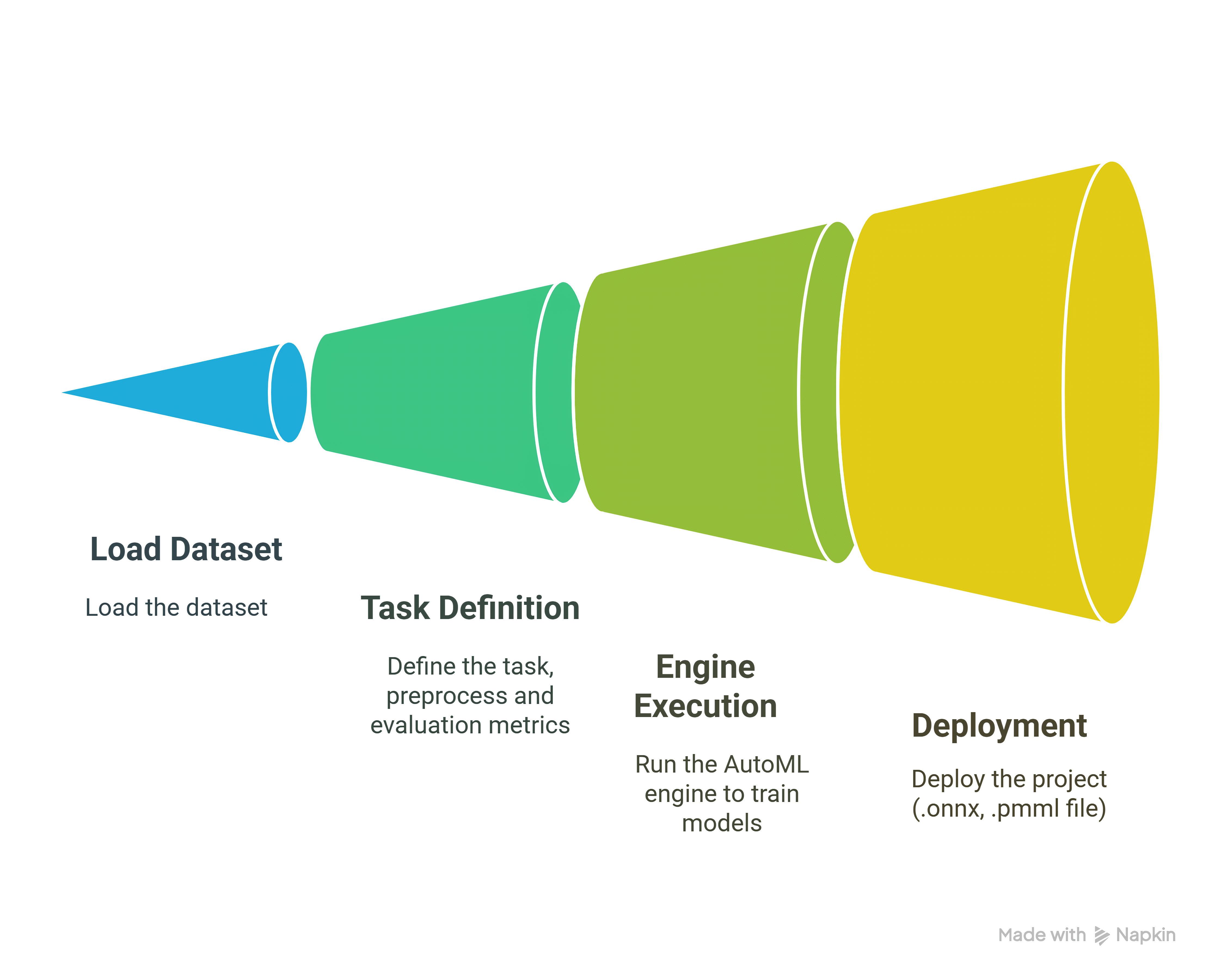}
    \caption{Overall Structure of proposed ZeroML Language.}
    \label{fig:tcanther}
\end{figure}

% Results
\section{Comparison with State-of-the-Art Languages}
The difference between the code style examples for (in order) Python, R and ZeroML will make clear that ZeroML is far ahead in simplicity, resourcefulness and provider development friendliness. Both R and Python are widely used languages in the field of data science, though both languages have lots of boilerplate code and explicit configurations in order to perform simple tasks of AutoML. For Pythons TPOT example, the users have to import lots of libraries, manually prepare the data with StandardScaler, split the data, initialize a TPOTClassifier with params and fit the model. In the same way, the R H2O snippet is doing the work of initializing the H2O engine, manually scaling the features, assigning the response and predictor variables, and also managing the training frame and seed for reproducibility.

ZeroML on the other hand compresses the entire AutoML process into a few clear and concise lines. The user only needs to load the dataset and call the automl function by providing input, target, preprocess, max\_time and evaluation. The technique is intuitive and does not require manual scaling, dataset splitting, or environment configuration. This level of abstraction is a huge win for writability and it still allows for the control of important configuration details. The model evaluation and reporting is also simplified via the report() method, which makes ZeroML very user-friendly, especially for those who are not proficient coders.

What's more, ZeroML's competitive execution time is achieved thanks to its compiled nature, giving it faster runtime performance and more efficient resource utilization as opposed to Python and R, which are interpreted languages. The lack of external library requirement also increases stability by bypassing common problems of package version and environment inconsistency. In summary, ZeroML is a high-productivity, low-code approach to AutoML that allows for fast prototyping and deployment with low cognitive overhead. These benefits make ZeroML an attractive option for practitioners and researchers aiming at efficient, maintainable and scaleable machine learning solutions.

\begin{figure}[!htpb]
    \centering
    \includegraphics[width=\linewidth]{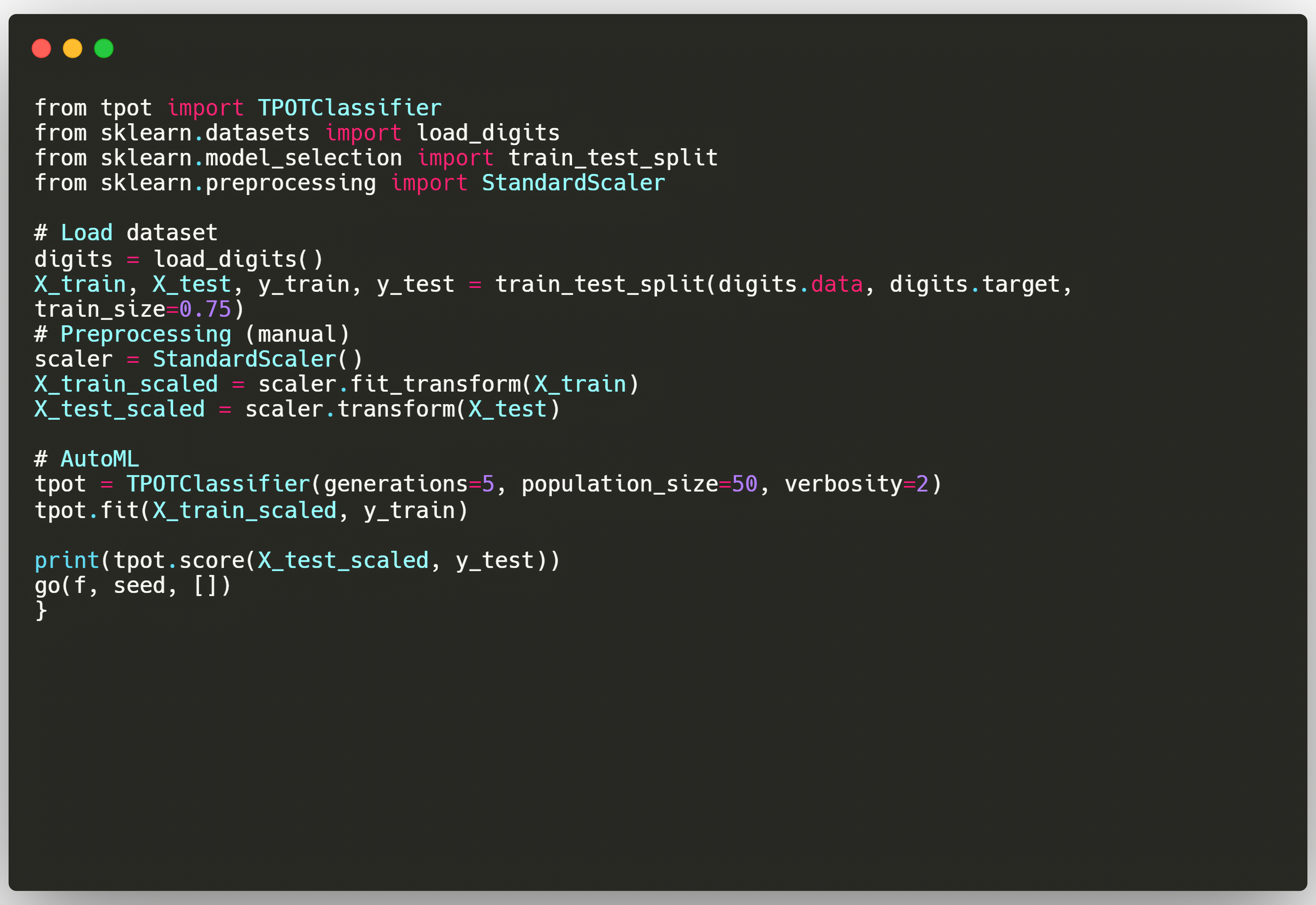}
    \caption{Code snippet with Python Language (TPOT).}
    \label{fig:tcanther}
\end{figure}

\begin{figure}[!htpb]
    \centering
    \includegraphics[width=\linewidth]{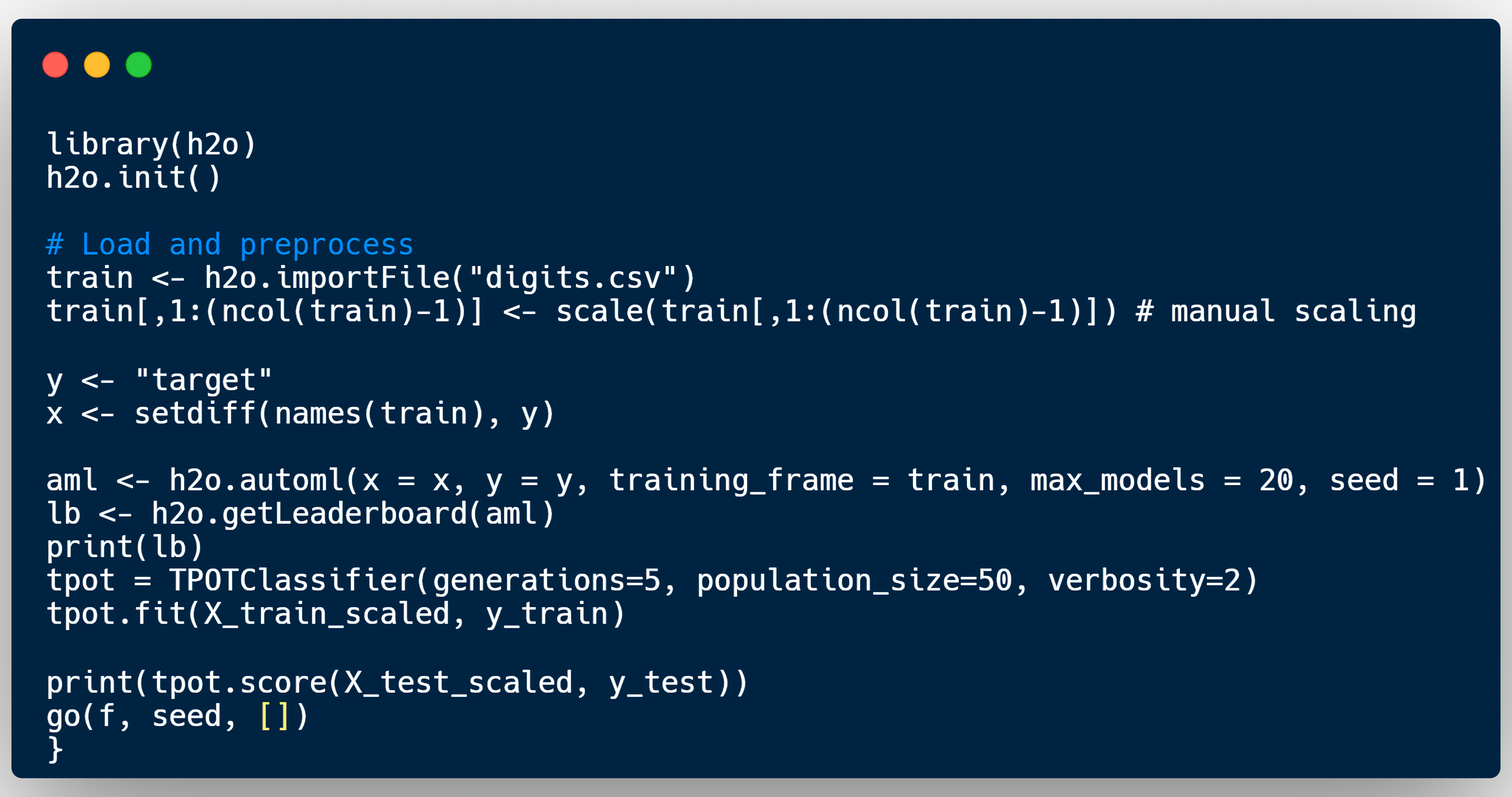}
    \caption{Code snippet with R Language (H2O).}
    \label{fig:tcanther}
\end{figure}

\begin{figure}[!htpb]
    \centering
    \includegraphics[width=\linewidth]{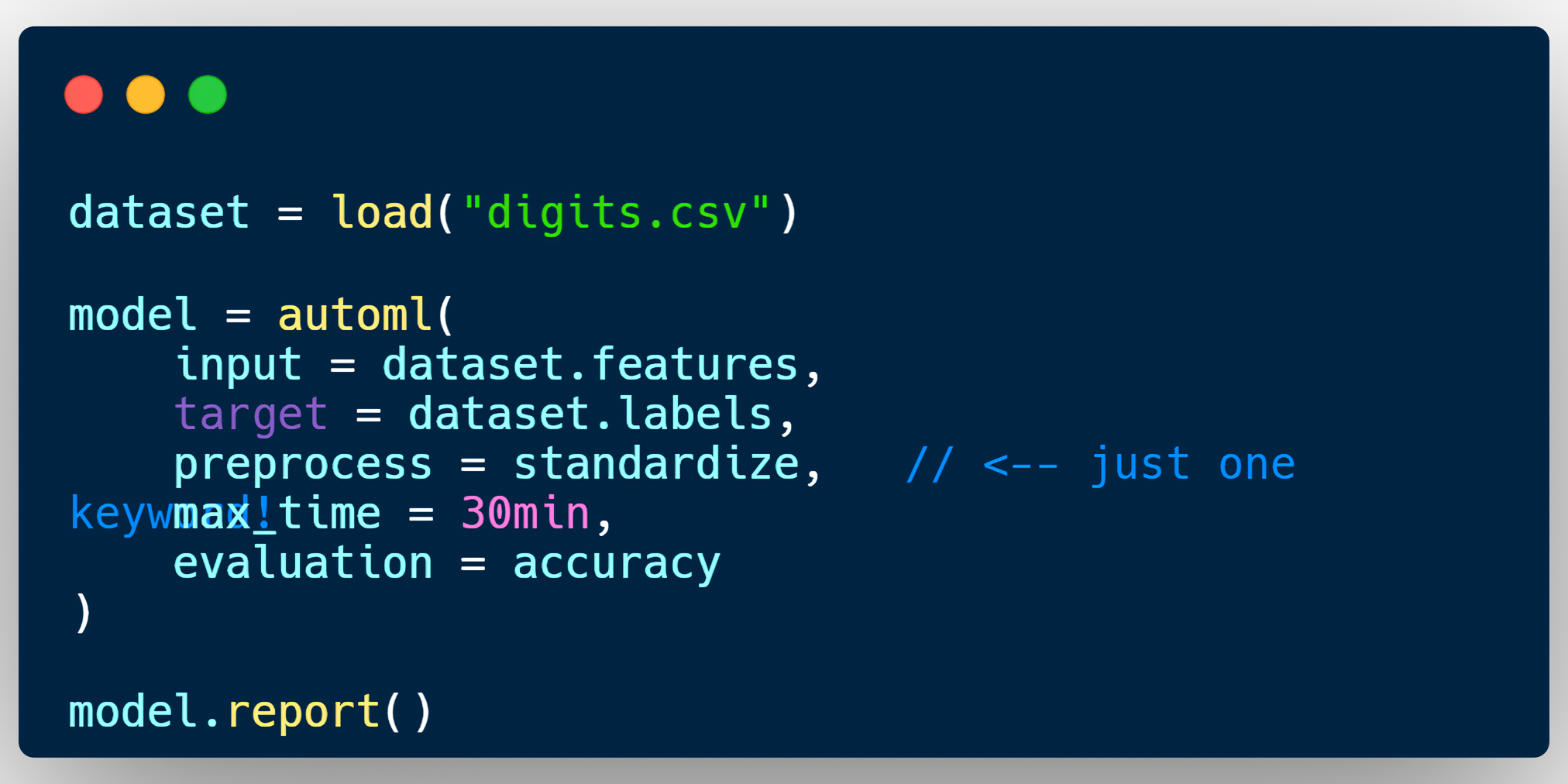}
    \caption{Code snippet with ZeroML Language.}
    \label{fig:tcanther}
\end{figure}
% Discussion
\section{High-Performance Execution and Scalable Design}

ZeroML is designed for maximizing the power of multithreading, ZeroML provides automatic concurrent data preprocessing, model training, and evaluation. Runtime architecture of torsimany is smart enough to handle thread pool and memory allocation well enough that it can use all the CPU cores without much of a configuration. This allows model search and opitmization to run in parallel, saving a significant amount of time for large datasets or intricate pipelines.

Scalability under ZeroML goes beyond raw performance. Its architecture enables local or distributed execution without any modification in design. ZeroML scales up or down to get the most out of your hardware, whether it’s a laptop or a cluster in the cloud. Additional memory optimization techniques such as lazy loading, in place transformations, and smart caching make it efficient and scalable for working with big data. In contrast with Python or R, where scaling typically means rewriting code for things like Spark or Dask, in ZeroML your syntax and workflow remain similarly brief. This means users – ranging from students to enterprise teams – can develop and scale models, without having to learn new tools or languages, and it’s the true AutoML solution that scales with the scope and complexity of the problem you are addressing.

\section{Inspiration and Consideration}

This language was designed mainly as an attempt to capture the expressiveness of high levels while avoiding a lot of the common object oriented traps. I wanted to take a middle ground between the understandability of functional programming and the abstraction capability of OO. More precisely, I deliberately set out to get rid of many of the popular pitfalls associated with OOP (like deep inheritance hierarchies, and excessive use of polymorphism) that make understanding program behaviour more difficult which only tended to lead to a headache later when you tried to debug it. Also inspired by the concept of having everything immutable by default but having one clearly defined method in each class which is allowed to change the state. A statically typed, expression based programming language that combines functional and a simplified object model. It's got composition over inheritance, immutability by default, and some explicit state mutation to better control side effects. It is designed for systems programming, with a focus on concurrency, productivity, and modularity so it can be used for large scale, production systems that need a high level of reliability.

\section{Syntax (Backus-Naur Form)}

<program> ::= { <statement> } \\

<statement> ::= <declaration> | <if-then-else> | <for-loop> | <function-call> | <expression> ";"\\

<declaration> ::= "let" <identifier> "=" <expression> ";"\\

<if-then-else> ::= "if" "(" <expression> ")" "{" <statement> "}" [ "else" "{" <statement> "}" ]\\

<for-loop> ::= "for" "(" <identifier> "in" <expression> ")" "{" <statement> "}"
\\
<function-call> ::= <identifier> "(" [ <expression> { "," <expression> } ] ")"\\

<expression> ::= <literal> | <identifier> | <function-call> | <binary-op>\\

<binary-op> ::= <expression> ("+" | "-" | "*" | "/") <expression>\\

% Conclusion
\section{Conclusion}

ZeroML is a step forward for automatic machine learning as it allows for a powerful multi-paradigm, compiled language to also be expressed in a simple, clean syntax with a modular structure. It makes the ML process easier, by providing a set of abstractions that allows to perform complex tasks and at the same time enables the use of highly performant models, while being scalable and of low runtime cost. Its low code overhead, multithreading support, and ready availability make it usable for your average software worker bee as well as the systems guy. By providing a smooth path to entry with rapid prototyping and deployment, ZeroML is both democratizing machine learning and raising the bar in terms of efficiency and accessibility for AutoML.

\section*{Acknowledgements}
This project was made possible through the insightful mentorship and academic guidance of Professor William Lord, whose expertise and encouragement played a pivotal role in shaping the vision and execution of ZeroML Language.

\section*{References}

\noindent
[1] Dissanayake, D., Navarathna, R., Ekanayake, P., \& Rajadurai, S. (2025). \textit{A survey of evaluating AutoML and automated feature engineering tools in modern data science}. Proceedings of ICEIS 2025, 218--225.

\noindent
[2] Feurer, M., Eggensperger, K., Falkner, S., Lindauer, M., \& Hutter, F. (2022). \textit{Auto-sklearn 2.0: Hands-free AutoML via meta-learning}. Journal of Machine Learning Research, 23(261), 1--61.

\noindent
[3] Li, Z., Zhou, X., Liu, W., Chen, L., \& Yang, F. (2021). \textit{FLAML: A Fast and Lightweight AutoML Library}. Microsoft Research.

\noindent
[4] Golovin, D., Solnik, B., Moitra, S., Kochanski, G., Karro, J., \& Sculley, D. (2017). \textit{Google Vizier: A service for black-box optimization}. In Proceedings of ACM KDD, 1487--1495.

\noindent
[5] Scott, I. A., et al. (2023). \textit{Evaluating automated machine learning platforms for use in healthcare}. Journal of Medical Internet Research, 25(7), e47521.

\noindent
[6] Bunay, P., Otero, I., Armada, S., et al. (2024). \textit{Easing the prediction of student dropout: Integration of AutoML and explainable AI}. Poster presented at the Educational Data Mining Conference, 2024.

\noindent
[7] Lindauer, M., et al. (2023). \textit{Review of the Year 2023 – AutoML Hannover}. Retrieved from \url{https://www.automl.org/blog/review-of-the-year-2023/}

\printbibliography

\end{document}